\documentclass{article}

\usepackage{PRIMEarxiv}

\usepackage[utf8]{inputenc} 
\usepackage[T1]{fontenc}    
\usepackage{hyperref}       
\usepackage{url}            
\usepackage{booktabs}       
\usepackage{amsfonts}       
\usepackage{nicefrac}       
\usepackage{microtype}      
\usepackage{lipsum}
\usepackage{tablefootnote}
\usepackage{color, colortbl}
\usepackage[misc]{ifsym} 
\usepackage{soul}
\usepackage{amsmath}
\usepackage{fixltx2e}
\usepackage{afterpage}
\usepackage{caption}
\usepackage{fancyhdr}       
\usepackage{graphicx}       
\graphicspath{{media/}}     

\definecolor{brown}{RGB}{229,225,224}
\definecolor{teal}{RGB}{178,216,216}

\usepackage{cite}
\usepackage{graphicx}
\usepackage[flushleft]{threeparttable}

\title{Influence of Medical Foreign Bodies on Dark-Field Chest Radiographs: First experiences
}

\author{
  \Letter \hspace{0.3em} Lennard Kaster$^{* 1, 2,3}$,  Henriette Klein$^{* 5}$, Alexander W. Marka$^\mathrm{{3}}$, Theresa Urban$^\mathrm{{1-3}}$, Sandra Karl$^\mathrm{{1-3}}$,\\
  \textbf{Florian T. Gassert$^\mathrm{{3}}$, Lisa Steinhelfer$^\mathrm{{3}}$, Marcus R. Makowski$^\mathrm{{3}}$, Daniela Pfeiffer$^\mathrm{{3,4}}$},
  \textbf{Franz Pfeiffer$^\mathrm{{1-4}}$} \\\\
  1 Chair of Biomedical Physics,  TUM School of Natural Sciences\\
  2 Munich Institute of Biomedical Engineering \\
  3 Department of Diagnostic and Interventional Radiology, Klinikum rechts der Isar\\
  4 Institute for Advanced Study \\
  5 Division of Thoracic Surgery, Klinikum rechts der Isar\\\\
  Technical University of Munich, Germany\\
  * Authors contributed equally to this work\\
  \Letter \hspace{0.3em} \texttt{lennard.kaster@tum.de} \\
}

\begin{document}
\maketitle

\section*{Clinical relevance statement}
Dark-field radiography reveals medical foreign bodies with reduced signals and fewer artifacts, enhancing overlay-free assessment of pulmonary tissue compared to conventional radiography.

\section*{Key points:}
\begin{enumerate}
  \item Dark-field chest radiographs showed reduced signals and artifacts compared to conventional radiography
  \item Highly attenuating foreign bodies can induce a beam-hardening induced dark-field signal
  \item Foreign bodies with a microstructure exhibited a positive true dark-field signal 
\end{enumerate}


\newpage
\section*{\center Declarations}
\textbf{Ethical approval and consent to participate}\\
Approval of the Institutional Review Board (Ethics Commission of the Medical Faculty, Technical University of Munich, Germany; reference no. 166/20S and 587/16 S) and the National Radiation Protection Agency was obtained prior to these studies. All patients gave written informed consent. \\\\
\textbf{Consent for publication}\\
Not applicable.\\\\
\textbf{Availability of data and materials}\\
Data generated or analyzed during this study are available from the corresponding author by reasonable request. \\\\
\textbf{Competing interests}\\
We report no conflict of interest.\\\\
\textbf{Funding}\\
Funded by the European Research Council (ERC, H2020, AdG 695045); the Deutsche Forschungsgemeinschaft (GRK 2274); the Federal Ministry of Education and Research (BMBF) and the Free State of Bavaria under the Excellence Strategy of the Federal Government and the Länder and the Technical University of Munich – Institute for Advanced Study. This work was carried out with the support of the Karlsruhe Nano Micro Facility (KNMF, www.kit.edu/knmf), a Helmholtz Research Infrastructure at Karlsruhe Institute of Technology (KIT)..\\\\
\textbf{Authors’ contributions}\\
All the authors contributed at the different stages of the study; 
LK and HK share first authorship as they contributed equally to this work. The literature review was carried out by LK and HK. LK, HK, AWM and DP curated and analyzed the data. TU, SK, FTG, LS, MRM, and DP contributed to the clinical methodology and the reader study. LK, HK, AWM, DP, FP investigated and interpreted the results. FP and DP conceptualized, administered and acquired funding for this study. All authors review the manuscript. All authors read and approved the final manuscript.\\\\

\newpage

\begin{abstract}
\textbf{Background:} Evaluating the effects and artifacts introduced by medical foreign bodies in clinical dark-field chest radiographs and assessing their influence on the evaluation of pulmonary tissue, compared to conventional radiographs.\\\\
\textbf{Methods:} This retrospective study analyzed data from subjects enrolled in clinical trials conducted between 2018 and 2021, focusing on chronic obstructive pulmonary disease (COPD) and COVID-19 patients. All patients obtained a radiograph using an in-house developed clinical prototype for grating-based dark-field chest radiography. The prototype simultaneously delivers a conventional and dark-field radiograph. Two radiologists independently assessed the clinical studies to identify patients with foreign bodies. Subsequently, an analysis was conducted on the effects and artifacts attributed to distinct foreign bodies and their impact on the assessment of pulmonary tissue.\\\\
\textbf{Results:} Overall, 30 subjects with foreign bodies were included in this study (mean age, 64 years $\pm$ 11 [standard deviation]; 15 men). Foreign bodies composed of materials lacking microstructure exhibited a diminished dark-field signal or no discernible signal. Foreign bodies with a microstructure, in our investigations the cementation of the kyphoplasty, produce a positive dark-field signal. Since most foreign bodies lack microstructural features, dark-field imaging revealed fewer signals and artifacts by foreign bodies compared to conventional radiographs.\\\\
\textbf{Conclusion:} 
Dark-field radiography enhances the assessment of pulmonary tissue with overlaying foreign bodies compared to conventional radiography. Reduced interfering signals result in fewer overlapping radiopaque artifacts within the investigated regions. This mitigates the impact on image quality and interpretability of the radiographs and the projection-related limitations of radiography compared to CT.\\\\
\textbf{Trial registration:} Retrospectively registered.
\end{abstract}

\section*{Introduction}
Lung diseases, such as chronic obstructive pulmonary disease, lung cancer, and pneumonia, pose significant challenges to global health \cite{Vos2016}. Early and accurate diagnosis is essential for the optimal management and treatment of these conditions. Radiologic imaging, particularly chest radiography, is a fundamental diagnostic tool due to its widespread availability, cost-effectiveness, and relatively low radiation exposure \cite{Schaefer-Prokop2002}. However, conventional radiography faces challenges in detecting early-stage lung diseases due to weak signals in the lung tissue and anatomical overlaps, especially in areas such as the ribs. \\
Dark-field radiography is a pioneering imaging technique that utilizes small-angle X-ray scattering at material interfaces within the sample under investigation to produce dark-field signals. Healthy pulmonary tissue generates a strong dark-field signal due to its numerous refracting tissue-air interfaces \cite{Gassert2021,Schleede2012}. Pathological pulmonary tissue resulting from destruction, inflammatory processes, or pulmonary masses exhibits a diminished dark-field signal due to a reduced number of alveolar air-tissue interfaces \cite{Hellbach2015,Scherer2017,Hellbach2018,Willer2021,Urban2022,Frank2022,Urban2023}. In contrast, surrounding structures such as osseous and soft tissues yield minimal dark-field signals \cite{Gassert2021}.
The medical applications of dark-field radiography were made possible by the technique introduced by Pfeiffer et al. in 2008 \cite{Pfeiffer2008}. Previous human studies on the first human dark-field chest radiography have shown the superiority of dark-field radiography in diagnosing and staging pulmonary diseases compared to conventional radiography \cite{Willer2021,Urban2022,Frank2022,Urban2023}.  \\
While the effects of foreign bodies on conventional radiographs have been investigated over decades, these still need to be explored for dark-field radiographs. Understanding the effects of foreign bodies in radiographs is crucial as they can obscure underlying pathology, complicate diagnosis, and lead to misinterpretation of radiographs \cite{Hogeweg2013}. It is essential to comprehend the influence of foreign bodies not only to ensure accurate diagnosis but also to leverage the complementary information provided by radiographs. Additionally, this understanding is crucial for developing and implementing effective correction algorithms for artifacts induced by foreign bodies. The main objective of this study is to outline the qualitative effects of medical foreign bodies on X-ray dark-field radiographs in human subjects.

\section*{Methods}
\subsection*{Subjects}
\begin{figure}[h]
\includegraphics[width = 0.8\textwidth]{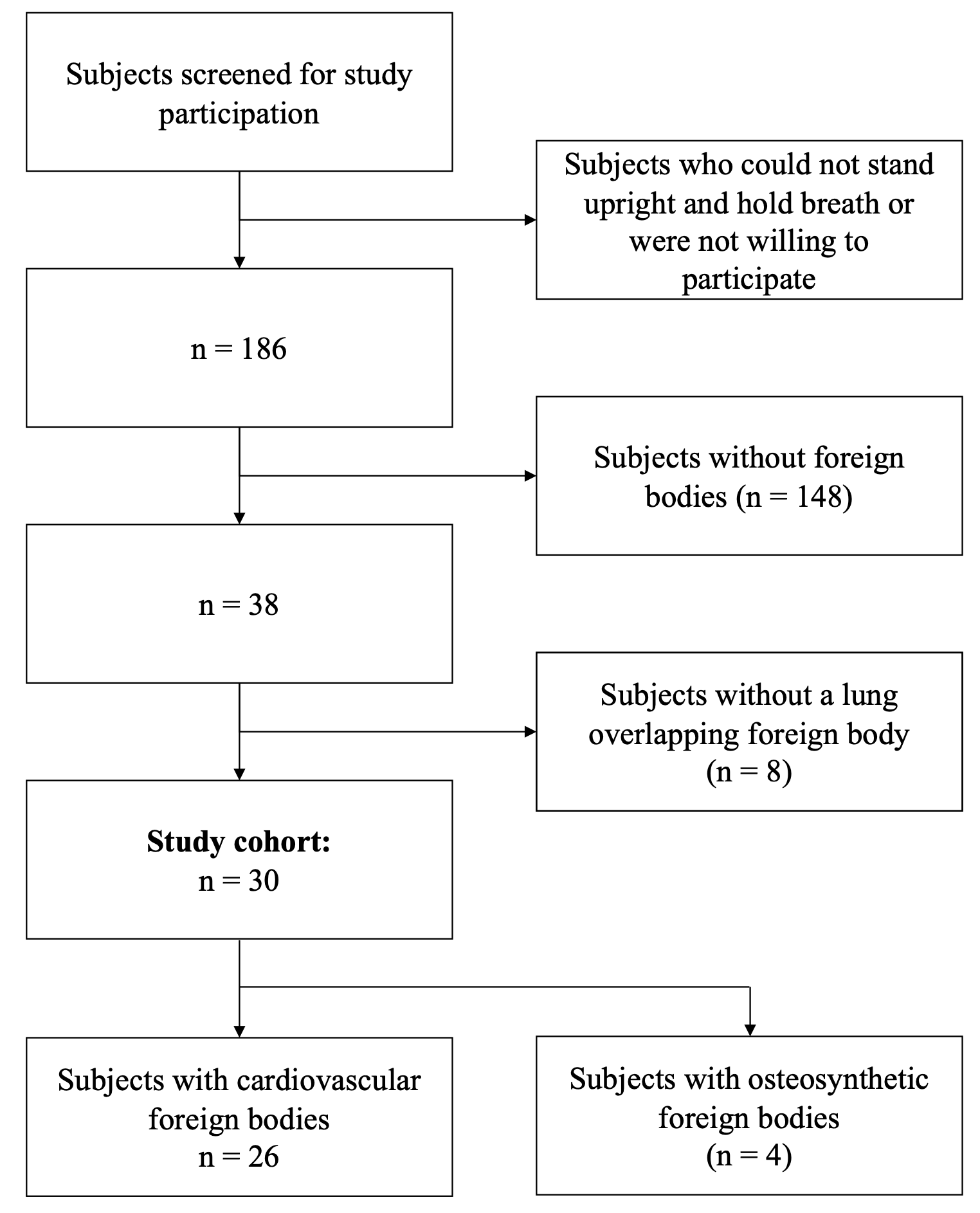}
\caption{Flowchart shows subject selection. Between 2018 and 2021, a total of 30 subjects were included in the study.}
\label{fig1}
\end{figure}

Approval of the Institutional Review Board (Ethics Commission of the Medical Faculty, Technical University of Munich, Germany; reference no. 166/20S and 587/16 S) and the National Radiation Protection Agency was obtained prior to these studies. The clinical studies are conducted in accordance with the Declaration of Helsinki (as revised in 2013).  All patients gave written informed consent. Between October 2018 and February 2021, participants at least 18 years old who underwent chest CT as part of their diagnostic work-up were screened for study participation. This retrospective study included participants who were capable of giving their consent and standing upright without any assistance. Exclusion criteria were the absence of foreign bodies and foreign bodies that didn’t overlap with the lung parenchyma in the radiographs. The selection process for this retrospective study can be found in \autoref{fig1}. 

\subsection*{Dark-field Chest X-Ray System}
Our study employed a clinical prototype for X-ray dark-field chest radiography. The system is based on a conventional X-ray system in combination with a 3-grating interferometer. The conventional X-ray system comprises an X-ray source (MRC 200 0508 ROT-GS 1003; Philips Medical Systems), collimators (R 302 DMLP DHHS; Ralco), and a flat-panel detector (PIXIUM 4343 F4; Trixel).
The interferometer introduces a periodic phase pattern onto the X-ray beam, resulting in an intensity pattern on the detector. Modifications of this intensity pattern can then be attributed to the subject’s attenuation, yielding the attenuation-based radiograph, like the conventional radiograph, and the amount of ultra-small-angle scattering, yielding the so-called dark-field radiograph \cite{Gassert2021}. Consequently, the dark-field and conventional radiographs can be acquired simultaneously with perfect temporal and spatial registration. The X-ray source is operated at 70~kVp with a tube filtration of 2.8~mm Al equivalent, and the duration of one image acquisition is about 7 seconds. The effective dose for the reference person is $35~\mu$Sv for the posterior-anterior examination and $45~\mu$Sv for the lateral \cite{Frank2022}. More detailed information on the setup and the reconstruction can be found in the literature \cite{Koehler2015,Kottler2007}.

\subsection*{Radiologic Analysis}
Two clinical radiologists (A.M., D.P.; 3,16 years of experience in radiography, of which 2,9 years in dark-field radiography) independently evaluated all images at our institution regarding the visible effects of foreign bodies in dark-field radiography. There was no disagreement between the radiologists. They identified patients with overlapping foreign bodies (n=30) and categorized them into two groups: cardiovascular and osteosynthetic foreign bodies. For all patients, the radiologists described the effects of the foreign bodies. From these 30 patients, 6 were selected as representative examples to illustrate the findings.

\section*{Results}

\subsection*{Patient Characteristics}

\begin{table}[h!]
\centering
\begin{threeparttable}
\caption{\small Patient Characteristics}
\label{tab:demographics}
\small
  \centering
    \begin{tabular}{lcccc}
    \midrule
    \rowcolor{teal}Parameter &  All \textit{(n=30)} &  Men \textit{(n=15)} & Women \textit{(n=15)}  \\
    \midrule
    Age (years) & 64 $\pm$ 11 & 66 $\pm$ 9 & 62 $\pm$ 13 \\
       \rowcolor{teal}Emphysema study & 25 & 14 & 11 \\
       COVID-19 study & 5 & 1 & 4  \\
    \bottomrule
    \end{tabular}
    \begin{tablenotes}
    \small
      \item Age is given as mean $\pm$ standard deviation.
    \end{tablenotes}
    \end{threeparttable}
\end{table}

The demographics of the patients are presented in \autoref{tab:demographics}. Dark-field and conventional radiographs of a total of 30 patients (mean age, 64 years $\pm$ 11 [standard deviation], 15 men) were included, which were acquired in our studies between 2018 and 2021. 
The images were originally obtained in the following studies: 25 were taken from our lung emphysema/COPD study, whereas 5 were used from the COVID-19 study. Among these 30 patients, 4 patients with osteosynthetic foreign bodies and 26 patients with cardiovascular foreign bodies were identified (Figure 1).
Of the 26 patients with cardiovascular foreign bodies, 24 had a Port-a-Cath, and 2 had an implantable cardioverter-defibrillator (ICD) unit. Among the 4 patients with osteosynthetic foreign bodies, 2 had dorsal stabilization, 1 had a sternal wire cerclage, and 1 had a clavicular plate.

\clearpage
\newpage

\subsection*{Radiologic Analysis}

\begin{figure}[h!]
\includegraphics[width=\textwidth]{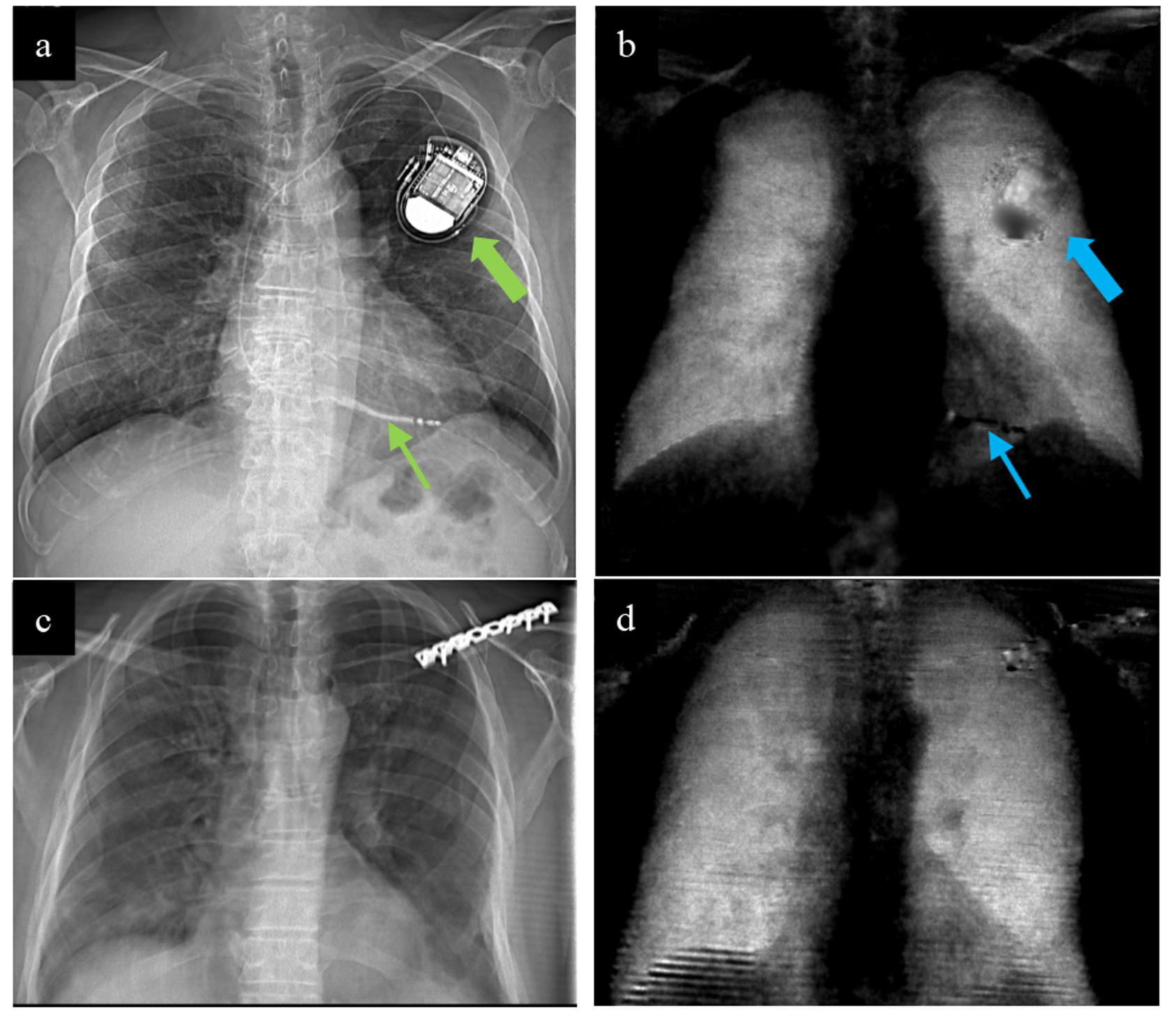}
\caption{\textbf{(a)} Attenuation-based and \textbf{(b)} dark-field radiographs of the thorax in a 70-year-old man. This patient has a one-lead implantable cardioverter-defibrillator (ICD) unit that projects infraclavicular onto the left upper chest (big arrows). The tip of the lead projects onto the right ventricle \textit{(small arrows)}. \textbf{(c)} Attenuation-based and \textbf{(d)} dark-field radiographs of the thorax in a 57-year-old man (COVID-study) with a plate osteosynthesis projecting onto the left clavicle. }
\label{fig2}
\end{figure}

\clearpage
\newpage

\begin{figure}[h!]
\includegraphics[width=\textwidth]{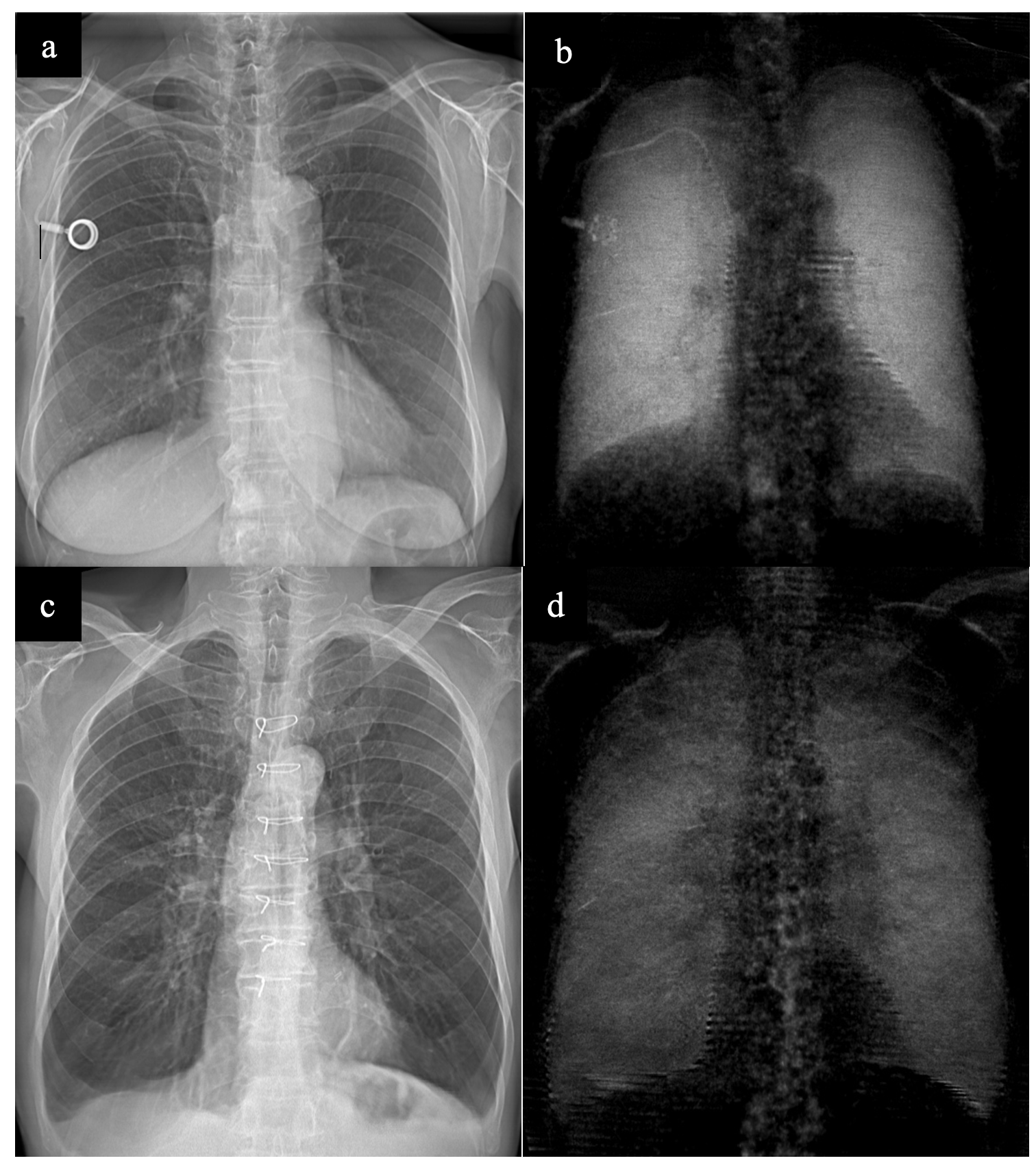}
\caption{\textbf{(a)} Attenuation-based and \textbf{(b)} dark-field radiographs of the thorax in a 71-year-old woman (COPD-study) with a Port-a-Cath-system that projects onto the right upper chest. \textbf{(c)} Attenuation-based and \textbf{(d)} dark-field radiographs of the thorax in a 71-year-old man (COPD-study) with seven sternal wire cerclages projecting onto the mediastinum. }
\label{fig3}
\end{figure}

\clearpage
\newpage

\begin{figure}[h!]
\includegraphics[width=\textwidth]{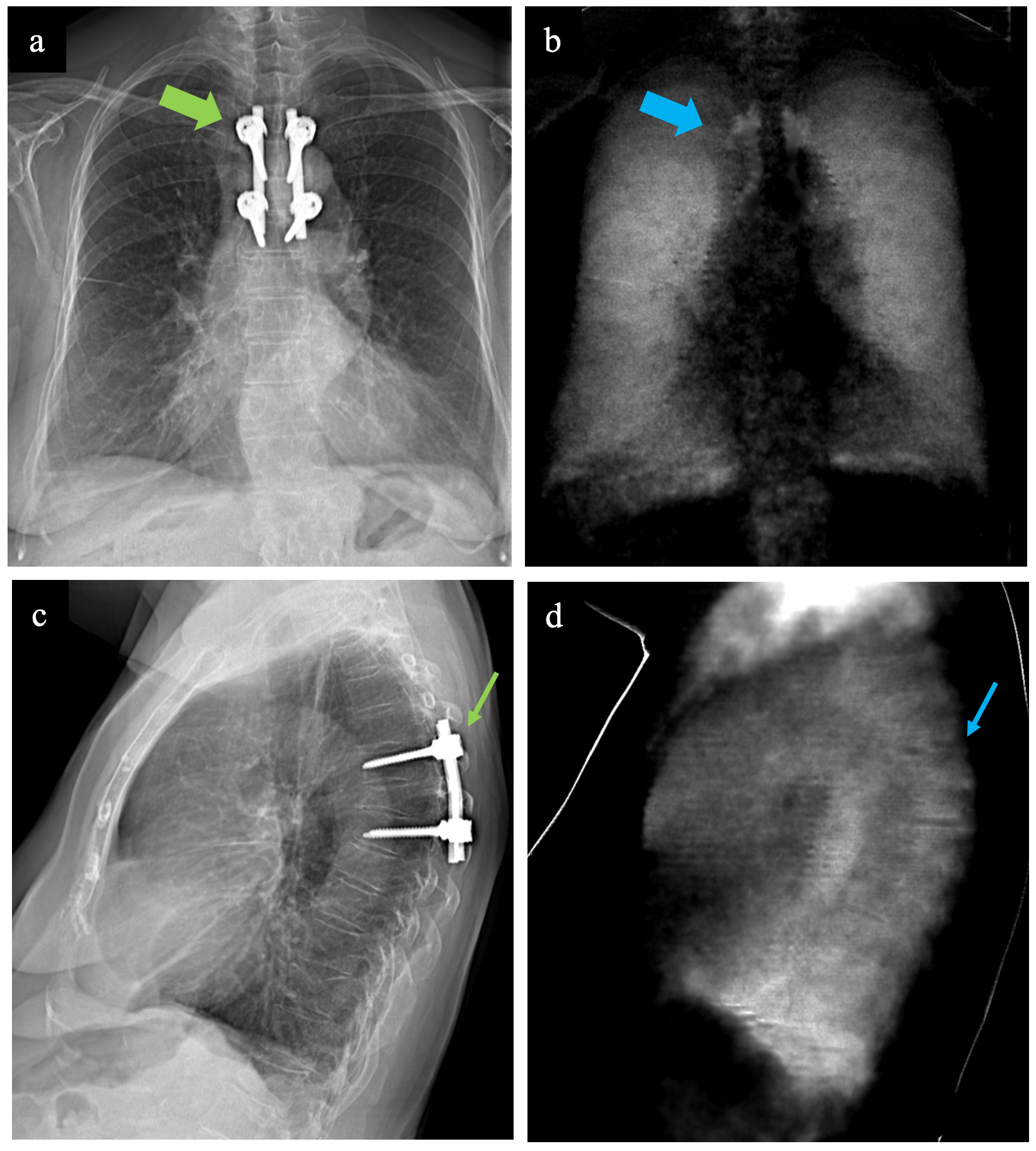}
\caption{\textbf{(a, c)} Attenuation-based images in AP-view \textbf{(a)} and lateral view \textbf{(c)}, complemented by \textbf{(b, d)} dark-field radiographs in AP-view \textbf{(b)} and lateral view \textbf{(d)} of the thorax in a 74-year-old woman (COPD-study). The patient has a dorstal stabilization \textit{(arrows)}. }
\label{fig4}
\end{figure}

\clearpage
\newpage

\begin{figure}[h!]
\includegraphics[width=\textwidth]{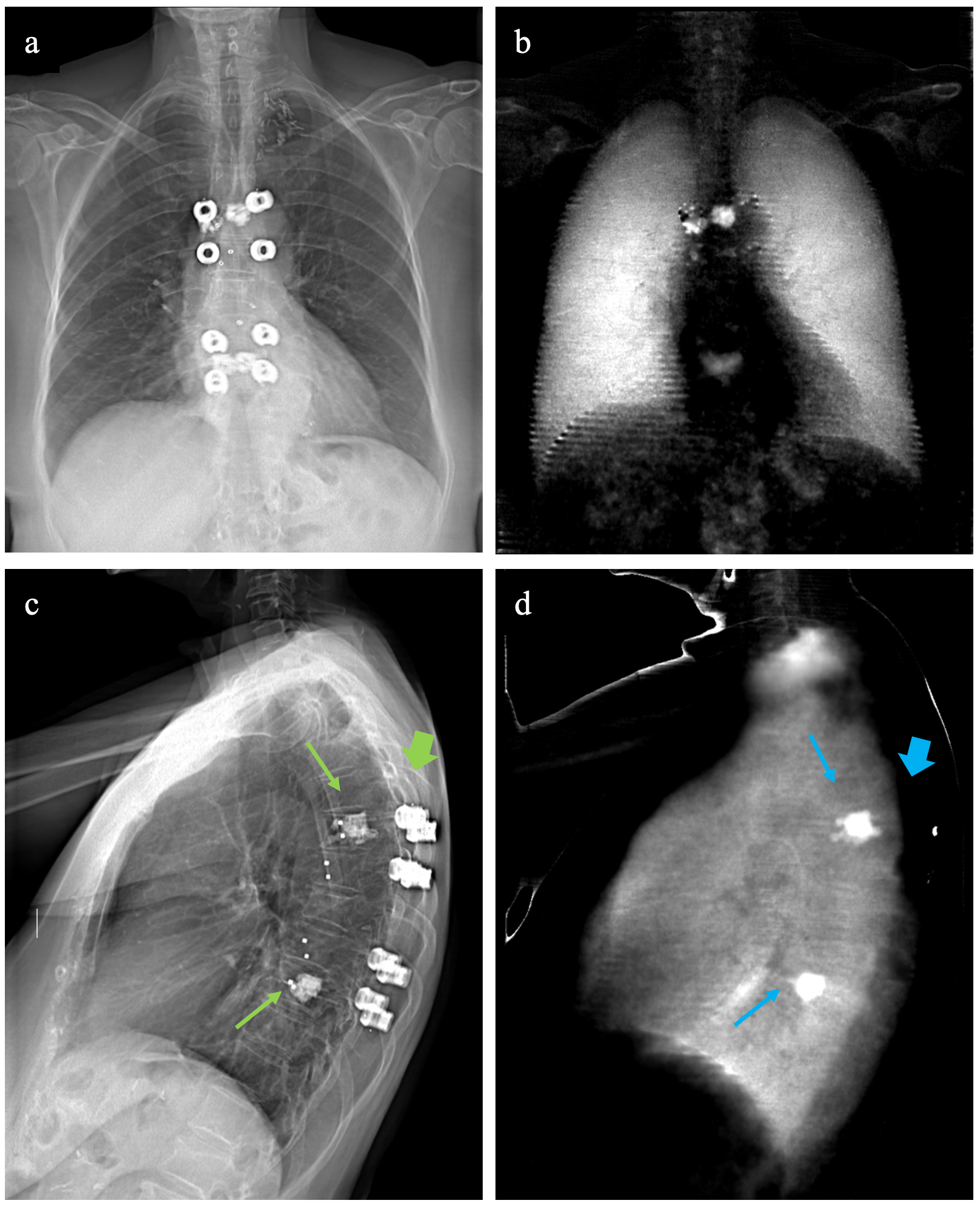}
\caption{\textbf{(a, c)} Attenuation-based radiographs in AP-view \textbf{(a)} and lateral view \textbf{(c)}, complemented by \textbf{(b, d)} dark-field radiographs in AP-view \textbf{(b)} and lateral view \textbf{(d)} of the thorax in an 80-year-old woman (COPD-study). The patient has a dorsal stabilization \textit{(large arrows)} and kyphoplasties \textit{(small arrows) }projecting onto the thoracic spine.}
\label{fig5}
\end{figure}

\clearpage
\newpage

The attenuation-based and dark-field radiographs of six representative subjects are shown (Figs. \ref{fig2}-\ref{fig5}).
In the case of a patient with a one-lead implantable cardioverter-defibrillator (ICD) (\autoref{fig2}), the ICD unit is distinctly radiopaque in the attenuation-based radiograph, projecting onto the left upper chest and overlaying the lung tissue. In the corresponding dark-field radiograph, the ICD unit generates a heterogeneous signal, characterized by regions of both negative and slightly positive dark-field contrast. Although lung tissue assessment is not feasible in conventional radiographs, the dark-field radiograph does not provide a reliable evaluation either, even with the reduction in overlapping signals. The lead tip appears radiopaque in the conventional radiograph and produces a negative signal in the dark-field radiograph. \\
A patient with plate osteosynthesis of the clavicle (\autoref{fig2}) exhibits a strong, radiopaque signal in the conventional radiograph, overlapping the left upper chest and obstructing the assessment of pulmonary tissue dorsal to the foreign body. Conversely, in the dark-field radiograph, the osteosynthesis plate produces only a marginal signal, with potential artifacts present. The plate appears with a pattern of alternating positive and negative signals in the dark-field radiograph, attributable to attenuation-dependent denoising. \\
An examination of a patient with a Port-a-Cath system (\autoref{fig3}) shows behavior similar to that of the clavicle osteosynthesis. In the attenuation-based radiograph, the port chamber is projected onto the lung tissue of the right upper chest with a positive signal. The dark-field radiograph depicts the port chamber with only slightly observable signals, but exhibiting an alternating pattern similar to that observed with the plate osteosynthesis, likely due to attenuation-dependent denoising. However, the associated tube induces only a diminutive signal in both the conventional and dark-field radiographs.
A cerclage on the spinal column (\autoref{fig3}) is solely visible in the attenuation-based radiograph, with minimal influence on the dark-field image. Notably, the cerclages do not overlap lung tissue and, therefore, do not affect the assessment of pulmonary tissue.  \\
The dorsal stabilization (\autoref{fig4}) appears radiopaque, overlaying the thoracic spine and parts of the lung in the attenuation-based radiograph. Although the overlapped area is only slightly in the anterior-posterior orientation, there is a notable overlap in the lateral orientation. In contrast, the dark-field radiograph reveals that the dorsal stabilization produces minimal dark-field signal in the posterior-anterior orientation and no discernible signal in the lateral orientation. \\
A patient with dorsal stabilization (\autoref{fig5}) exhibits a robust radiopaque signal in the attenuation-based radiograph for both posterior-anterior and lateral orientations. However, no signal is detectable in the lateral dark-field image. In the posterior-anterior orientation, a similar behavior as observed with the osteosynthetic plate is apparent. An alternating pattern of positive and negative dark-field signals, attributable to denoising, is visible in the upper parts of the dorsal stabilizations. The lower dorsal stabilizations exhibit slightly weaker signals in the attenuation radiograph compared to the upper parts, which is also noticeable in the dark-field radiograph where the alternating pattern is absent. In contrast, the cementation of the kyphoplasty is visible in the conventional radiographs and produces strong signals in the lateral dark-field image.

\section*{Discussion}
This study aimed to evaluate the impact of medical foreign bodies on dark-field chest radiography compared to conventional radiographs and its potential implications for image interpretation by radiologists. Unlike conventional radiography, which relies on absorption, dark-field radiography is based on ultra-small-angle scattering. The influence of foreign bodies on dark-field imaging is more complex than in standard attenuation-based radiography, where only material density and corresponding absorption are relevant for image generation. In dark-field radiography, the material composition and microstructure of foreign bodies, especially the material interfaces within them, must be considered. Our findings demonstrate that medical foreign bodies can impact the diagnostic value of dark-field radiographs for regions in its beam path, depending on the foreign body. However, medical foreign bodies result in less image interference in dark-field radiographs compared to conventional radiographs. Specifically, the most common foreign body, the Port-a-Cath \cite{MacMahon1991}, made of metal (platinum, titanium), POM (polyoxymethylene), or both \cite{Guiffant2017,Cowley2011}, has a minimal impact. Also, for osteosynthetic screws and plates, it was confirmed that the signal in dark-field radiographs is only slightly positive, whereas strong overlapping signals are observed in conventional radiographs. The reduced signals and artifacts in the dark-field radiograph, as compared to the attenuation-based conventional radiograph, can be attributed to the fact that most medical foreign bodies are made of materials that strongly attenuate X-rays but lack fine microstructures. This applies not only to the Port-a-Caths, but also to the pacemakers \cite{Cowley2011,Holmes2007,Perez-Nicoli2018,Mallela2004} and spinal fixations \cite{Cowley2011}. In contrast, lung tissue only weakly attenuates X-rays and comprises millions of alveoli that strongly scatter X-rays, thus producing a strong dark-field signal and weakly attenuation signal. Our evaluation indicates that dark-field radiographs improve the assessability of lung tissue in the presence of medical foreign bodies. While most foreign bodies appear radiopaque in conventional radiographs, they produce minimal to no dark-field signal. A distinct advantage of the dark-field chest radiography system is its ability to acquire perfectly temporally and spatially registered conventional and dark-field chest radiographs. This simultaneous acquisition provides two independent sets of complementary radiographs about foreign bodies and the subject being studied. The capability is advantageous when diagnosing lung diseases, as the complementary information from the two radiographs can be utilized even if foreign bodies significantly overlap the lung in one radiograph. Additionally, the precise assessment of foreign body position is feasible for both strongly attenuating foreign bodies in the conventional radiograph and weakly attenuating foreign bodies with microstructure in the dark-field radiograph, as seen in the cementation of dorsal stabilization. However, many medical devices are made of platinum so that the position can be precisely determined during the procedure \cite{Cowley2011}. The dark-field radiograph could then be used to make a diagnosis in subsequent X-ray examinations thanks to the improved assessment of the lung tissue. This complementary information might enhance diagnostic accuracy, enabling clinicians to leverage the distinct advantages of both imaging modalities to more effectively identify and evaluate both the lung tissue as well as foreign bodies.\\

However, for pacemakers composed of highly attenuating metals, signals were observed in dark-field radiographs despite the absence of a microstructure. These artifacts, located at the pacemaker site, can be attributed to the beam-hardening-induced dark-field signal. The beam-hardening induced dark-field signal has already been investigated and demonstrated in previous non-clinical experimental studies \cite{Yashiro2015}. Beam-hardening, a well-known effect in X-ray imaging, occurs due to the energy-dependent attenuation of the sample, which hardens the polychromatic X-ray spectrum. In dark-field radiography, a hardened X-ray spectrum changes the intensity pattern, which is then misinterpreted as a dark-field signal \cite{Yashiro2010}. For pacemakers, the dark-field signal from tissue along the same beam path cannot be reliably evaluated, although the induced signals are lower than in the conventional radiograph. Correcting for the beam-hardening-induced dark-field signal is challenging due to its interaction with attenuation and is part of ongoing research. The current correction method assumes a fixed ratio of attenuation caused by soft tissue and bone within the patient. Given that the relative contributions based on the patient’s body composition are unknown, this correction assumes equal contributions of soft tissue and bone \cite{Frank2022}. This approximation is rough and becomes even less accurate in the presence of a metallic foreign body. Consequently, the current correction method cannot adequately correct for beam-hardening-induced dark-field signals in the presence of foreign bodies. Future correction algorithms may involve simulating induced artifacts \cite{Pelzer2016}, using models to characterize the dark-field signal \cite{Taphorn2023}, or determining body composition on a pixel-by-pixel basis using material decomposition algorithms based on spectral information from spectral detectors (dual-layer detectors) or spectral acquisitions (kVp-switching, sequential scanning). \\

While the study provides first experiences into the qualitative analysis of foreign bodies in clinical dark-field chest radiography, limitations exist. We did not incorporate potential diagnosis of patients in areas affected by foreign bodies, which limits the clinical applicability. Additionally, the evaluation of the effects and artifacts induced by foreign bodies on dark-field chest radiographs was only qualitative and not quantitative. The study focuses on patients with diseases (COVID-19 and COPD) that induce global signal changes \cite{Urban2022,Frank2022}. To overcome these limitations, future studies should include lung diseases that lead to local signal changes in the areas overlapped by the foreign body to visualize the impact of the lower overlapping signals and artifacts in dark-field chest radiographs compared to conventional radiographs. Another advancement for future studies should develop quantitative measures for signal and artifact evaluation in dark-field chest radiographs. This would provide a more objective assessment of the impact of foreign bodies on image quality. A potential approach to overcome these challenges could involve phantom studies, that allow for a precise placement of foreign bodies in specific regions of the lung. Additionally, phantom studies would facilitate quantitative comparisons, which are challenging in human patients due to the inherent variability of lung tissue among individuals. \\

In conclusion, this study provides the first insights into the effects of foreign bodies on clinical dark-field chest radiographs. Our findings demonstrate a key advantage of the clinical dark-field system: the simultaneous acquisition of perfectly temporally and spatially registered conventional and dark-field radiographs. This unique feature provides complementary information that can be leveraged to enable precise localization of foreign bodies while enhancing the assessment of pulmonary tissue. Dark-field chest radiographs showed potential to enhance the overlay-free assessment of pulmonary tissue in 2D radiography in the presence of foreign bodies.



\end{document}